\documentstyle[prd,aps]{revtex}

\newcommand{\bq}{\begin{equation}}
\newcommand{\eq}{\end{equation}}
\newcommand{\bqn}{\begin{eqnarray}}
\newcommand{\eqn}{\end{eqnarray}}
\newcommand{\nb}{\nonumber}
\newcommand{\lb}{\label} 
 
\begin{document}
 

\title { 
On the back reaction of gravitational and particle emission
and  absorption from straight thick cosmic strings: A toy model}
\author{ M.L. Bedran\thanks{bedran@if.ufrj.br} and E.M. de S\'a }
\address{ Instituto de F\' {\i}sica, Universidade Federal 
do Rio de Janeiro,
Caixa Postal 68528, Cep 21945-970, Rio de Janeiro~--~RJ, Brazil}
\author{Anzhong Wang\thanks{wang@dft.if.uerj.br}}
\address{Departamento de F\' {\i}sica Te\' orica,
Universidade do Estado do Rio de Janeiro,
Rua S\~ ao Francisco Xavier 524, Maracan\~a,
20550-013 Rio de Janeiro~--~RJ, Brazil}
\author{Yumei Wu \thanks{e-mail address: yumei@dmm.im.ufrj.br}}
\address{Instituto de Matem\'atica, Universidade Federal do Rio
de Janeiro, Caixa Postal $68530$, Cep. $21945-970$,
Rio de Janeiro~--~RJ, Brazil}


\maketitle

\begin{abstract}

The emission and absorption of gravitational waves and massless 
particles of an infinitely long  straight cosmic string with finite 
thickness are studied. It is shown in a general term
that the back reaction of the emission and absorption {\em always} makes
the symmetry axis of the string singular. The singularity is a scalar
singularity and cannot be removed. 

\end{abstract}



 \vspace{.6cm}

\section{INTRODUCTION}
 
Topological defects that may have been formed in the early Universe
have been studied extensively since the pioneering work of Kibble
\cite{Kibble1976}. This is mainly due to their implications to the
formation of the large-scale structure of the Universe and the formation 
of galaxies \cite{VS1994}. During the past twenty years, the scenario of the
structure and galaxies formations from topological defects has experienced 
several revolutionary stages \cite{VS1994}. In particular, it was shown 
in 1997 \cite{PST1997} that the models were seriously in conflict with the 
observational data from COBE, while a year later \cite{BRA1998} it was
found that, after adding the cosmological constant into the models and by 
properly tuning some relevant parameters, consistent predictions  
can be achieved.

In this Letter, we shall not be concerned with the above mentioned 
problem, but stress another aspect that has been ignored in most numerical 
simulations \cite{VS1994}, that is, the back reaction of gravitational 
waves and particles emitted by the defects. As shown in \cite{SB1988}
by studying the full coupled Einstein-Maxwell-Higgs equations, 
cosmic strings always emit gravitational waves and particles due to their
contractions. The radiation continues until the strings settle into their
static configurations. 

The study of the back reaction of the radiation
 is not trivial \cite{VS1994}.
To have it attackable, in this Letter we shall consider a 
very ideal situation, that is, an infinitely long straight cosmic string 
with its energy-momentum tensor (EMT) being given by
\bq
\lb{eq1}
T_{\mu\nu}^{cs} = \sigma(t, r)\left\{u_{\mu}u_{\nu} - 
z_{\mu}z_{\nu}\right\},
\eq
where $u^{\lambda}u_{\lambda} = - z^{\lambda}z_{\lambda} = + 1$. It may
be argued that the model is too artificial to have anything to do with a
realistic cosmic string. However, we believe that it does give
some hint on how an important role the back reaction of gravitational 
waves and particles emitted by cosmic strings may play. Moreover,
it has been shown by several authors
 that the EMT given by Eq.(\ref{eq1}) for a straight
cosmic string is valid at least to the first-order approximation 
\cite{VS1994}. Therefore, the results obtained  
 here are expected to be valid to the  same order, too.  

The gravitational and particle radiation from an infinitely long thick  
cosmic string was studied recently by Wang and Santos, and they found 
that the back reaction of the radiation 
{\em always} turns the symmetry axis of the string into a 
spacetime singularity \cite{WS1996}. 
In the above considerations, cosmic strings were assumed only emitting
gravitational and particle radiation. However, since the spacetime
outside of the string is curved,  the outgoing radiation is always
expected to be backscattered by the spacetime curvature, so the ingoing
radiation in general also exists \footnote{It should be noted  
that the backscattering argument for the existence of the ingoing
radiation is not superficial. In fact, in black hole
physics it is exactly this argument that leads to the existence of
both ingoing and outgoing radiations, and the interaction of them 
is the crucial point that turns
Cauchy horizons into spacetime singularities and whereby the 
predictability of the Einstein theory is preserved. For the details
on this aspect, we refer readers to \cite{PI1990}.}. 
Yet, the useful cosmic strings usually are assumed to be formed 
at the temperature $T \approx 10^{15} 
Gev$. With such a high temperature, it would be expected that the Universe 
was filled with various kinds of radiation, gravitational and particle-like, 
with very high velocities. Such, as a first-order approximation, we may 
consider them as null dust. Therefore, for a more realistic model 
both outgoing and ingoing radiations
should be included. Then, a natural question is: What roles do the 
interaction between the ingoing radiation and cosmic string and the 
interaction between the ingoing and outgoing radiations play?
Using the general results obtained by Letelier
and Wang \cite{LW1994}, we shall study these interactions. In particular,
we shall show that the non-linear interaction between the ingoing
and outgoing radiations makes the singularity on the symmetry axis 
stronger.

\section{ THE EMISSION AND ABSORPTION OF COSMIC STRINGS}

To start with, let us consider the  metric for the spacetimes with
cylindrical symmetry \cite{Th1965}
\bq
\lb{eq2}
 ds^2 = e^{2(K-U)}(dt^{2} - dr^{2}) 
- e^{2U}dz^{2} - e^{-2U}W^{2}d\varphi^{2},
\eq
where $K, U,$ and $W$ are functions of t and r only, and $\{t, r, z,
\varphi\}$ are the usual cylindrical coordinates. The existence of an
axis  and the local-flatness condition at the axis require
\bq
\lb{eq3}
|\partial\varphi|^{2} = - g_{\varphi\varphi} 
= e^{-2U}W^{2} \rightarrow O(r^{2}), 
\eq
as $ r \rightarrow 0$, where we have taken $r = 0$ as the location
of the axis. Taking Eq.(\ref{eq1}) as the source of the
Einstein field equations $ G_{\mu\nu} = T^{cs}_{\mu\nu}$, Shaver and Lake
\cite{SL1989} showed that the metric coefficient $K$ is a function of
$t$ only, i.e., $K(t, r) = K^{cs}(t)$, and that
\bq
\lb{eq4}
\sigma = e^{2(U - K)}\frac{W_{,tt} - W_{,rr}}{W},
\eq
where $(\;)_{,x} \equiv \partial(\;)/\partial x$, and
\bq
\lb{eq5}
u_{\mu} = (g_{tt})^{1/2}\delta^{t}_{\mu},\;\;\;
z_{\mu} = (- g_{zz})^{1/2}\delta^{z}_{\mu}.
\eq
If we assume that  the energy density of the string is finite as
$r \rightarrow 0$, it can be shown that this condition together with
Eq.(\ref{eq3}) require \cite{WS1996}
\bqn
\lb{eq6}
W(t,r) &\sim& w_{1}(t)r + w_{3}(t)r^{3} + O(r^{4}),\nb\\
 e^{U(t,r)} &\sim& w_{1}(t) + a_{1}(t)r + O(r^{2}), 
\eqn
as $r \rightarrow 0$. Eq.(\ref{eq6}) and
$K = K^{cs}(t)$ are sufficient to ensure that  the axis
is free of spacetime singularities.
To show this, let us first choose a null tetrad,
\bqn
\lb{eq7}
L^{\mu} &=& \frac{e^{U-K}}{\sqrt{2}}
(\delta^{\mu}_{t} - \delta^{\mu}_{r}),\;\;\;
N^{\mu} = \frac{e^{U-K}}{\sqrt{2}} 
(\delta^{\mu}_{t} + \delta^{\mu}_{r}),\nb\\
M^{\mu} &=& \frac{1}{\sqrt{2}}(e^{- U}\delta^{\mu}_{z} + 
\frac{i}{W}\delta^{\mu}_{\varphi}),\;\;\;
\bar{M}^{\mu} = \frac{1}{\sqrt{2}}(e^{- U}\delta^{\mu}_{z} - 
\frac{i}{W}\delta^{\mu}_{\varphi}).
\eqn
Then, it can be shown that the non-vanishing components of the Ricci
tensor $R_{\mu\nu}$ and the Weyl tensor $C_{\mu\nu\lambda\delta}$ are
given by
\bqn
\lb{eq8}
\Phi_{00} &=& \frac{1}{2}S_{\mu\nu}L^{\mu}L^{\nu} = 
\frac{1}{4}\sigma,\;\;\;
\Phi_{02} = \frac{1}{2}S_{\mu\nu}M^{\mu}M^{\nu} = 
- \frac{1}{4}\sigma,\nb\\
\Phi_{22} &=& \frac{1}{2}S_{\mu\nu}N^{\mu}N^{\nu} =  
\frac{1}{4}\sigma,\;\;\;
\Lambda = - \frac{1}{24} R = \frac{1}{12}\sigma,\\
\lb{eq9}
\Psi_{0} &=& - C_{\mu\nu\lambda\delta}L^{\mu}M^{\nu}
L^{\lambda}M^{\delta}\nb\\
&=& - \frac{1}{4}\sigma - \frac{1}{2}e^{2(U-K)}\left[
U_{,tt} + U_{,rr} - 2U_{,tr} \right. \nb\\
& & \left. + 3(U_{,t} - U_{,r})^{2} 
 - 2K_{,t}(U_{,t} - U_{,r})\right],\nb\\
\Psi_{2} &=& - \frac{1}{2}C_{\mu\nu\lambda\delta}
\left[L^{\mu}N^{\nu}L^{\lambda}N^{\delta} -
L^{\mu}N^{\nu}M^{\lambda}\bar{M}^{\delta}\right]\nb\\
&=&  \frac{1}{12}\sigma + \frac{1}{2}e^{2(U-K)}\left[
U_{,tt} - U_{,rr} + U_{,t}^{2} - U_{,r}^{2}\right],\nb\\
\Psi_{4} &=& - C_{\mu\nu\lambda\delta}N^{\mu}\bar{M}^{\nu}
N^{\lambda}\bar{M}^{\delta}\nb\\
&=& - \frac{1}{4}\sigma - \frac{1}{2}e^{2(U-K)}\left[
U_{,tt} + U_{,rr} + 2U_{,tr} \right.\nb\\
& & \left. + 3(U_{,t} + 
U_{,r})^{2} - 2K_{,t}(U_{,t} + U_{,r})\right],
\eqn
where $\sigma$ is given by Eq.(\ref{eq4}), and $S_{\mu\nu} \equiv
R_{\mu\nu} - g_{\mu\nu}R/4$. From Eqs.(\ref{eq4})-(\ref{eq9}) we can see
that all the non-vanishing components, $\Phi_{AB}$ and $\Psi_{A}$, are
regular at the axis. On the other hand, the fourteen scalars built from
the Riemann tensor are the combinations of these quantities
\cite{CW1977}. From there we can see that if these components are not
singular at the axis, so do the fourteen scalars. This completes the
proof to the above claim.

To consider the emission and absorption of gravitational waves and null
dust fluids, from Eq.(2.9) of \cite{LW1994}, we have the following:
considering the outgoing and ingoing null dust fluids with the EMT's 
given by
\bq
\lb{eq10}
T^{out}_{\mu\nu} = \rho^{out} N_{\mu}N_{\nu},\;\;\;
T^{in}_{\mu\nu} = \rho^{in} L_{\mu}L_{\nu},
\eq
we find that these null dust fluids have only contribution 
to the metric coefficients $g_{tt}$ and $g_{rr}$,
where the two null vectors $L^{\mu}$ and $N^{\mu}$ 
given by Eq.(\ref{eq7}) 
define, respectively, the ingoing and outgoing null geodesic
congruence. In particular, if 
\bq
\lb{eq10a}
\{K^{cs}(t, r), U^{cs}(t, r), W^{cs}(t, r)\}
\eq
 is a solution of the Einstein field equations $G_{\mu\nu}^{cs} =
T^{cs}_{\mu\nu}$, then
\bq
\lb{eq11}
\left\{K, U, W\right\} = 
\left\{K^{cs} + a(u) + b(v), U^{cs}, W^{cs}\right\},
\eq
is a solution of the Einstein field equations
\bq
\lb{eq12}
R_{\mu\nu} - \frac{1}{2}g_{\mu\nu}R  = 
T^{cs}_{\mu\nu} + T^{out}_{\mu\nu}
+ T^{in}_{\mu\nu},
\eq
where $T^{cs}_{\mu\nu},\; T^{out}_{\mu\nu}$ and  $T^{in}_{\mu\nu}$ are
defined, respectively, by Eqs. (\ref{eq1}) and (\ref{eq10}), with
\bqn
\lb{eq13}
\rho^{in} &=& - \frac{a'(u)(W_{,t} + W_{,r})}{\sqrt{2} W},\;\;\;
\rho^{out} = - \frac{b'(v)(W_{,t} - W_{,r})}{\sqrt{2} W},\nb\\
\sigma &=& e^{2(U-K^{cs})+ a(u) + b(v)}\frac{(W_{,tt} - W_{,rr})}{W}, 
\eqn
where $a(u)$ and $b(v)$ are arbitrary functions, $u \equiv (t +
r)/\sqrt{2},\; v \equiv (t - r)/\sqrt{2}$, and a prime denotes the
ordinary derivative with respect to the indicated argument.
Eq.(\ref{eq12}) shows that the solution of Eq.(\ref{eq11}) indeed
represents a cosmic string accompanied by two null dust fluids, one is
outgoing, and the other is ingoing. Note that it is always possible by
properly choosing the two functions $a(u)$ and $b(v)$ to ensure the
energy densities of the two null fluids to be non-negative. In the
following, we always assume that this is the case. To show that the
outgoing null dust fluid is emitted by the string, let us consider the
conservation law $T_{\mu\nu;\lambda}g^{\nu\lambda} = 0$, which can be
written in the form
\bq
\lb{eq14}
T^{cs}_{\mu\nu;\lambda}g^{\nu\lambda} = J^{out}_{\mu} + J^{in}_{\mu},
\eq
where
\bqn
\lb{eq15}
J^{out}_{\mu} &\equiv& - \left(\rho^{out}
N_{\mu}N_{\nu}\right)_{;\lambda}g^{\nu\lambda}=
\frac{1}{2}\sigma b'(v)N_{\mu},\nb\\
J^{in}_{\mu} &\equiv& - \left(\rho^{in}
L_{\mu}L_{\nu}\right)_{;\lambda}g^{\nu\lambda}=
\frac{1}{2}\sigma a'(u)L_{\mu}.
\eqn
These expressions show clearly that the outgoing null dust fluid is
indeed emitted by the string, while the ingoing one is absorbed by it.

To obtain a cosmic string with a finite thickness, we may follow
\cite{WS1996} to cut the spacetime along a hypersurface, say, $r
= r_{0}(t)$, and then join the part with $r \le r_{0}(t)$ with a
spacetime where only the two null fluids exist. However, since here we
are mainly interested in the properties of the spacetime near the axis,
we shall not consider such a junction, and simply assume, without loss
of generality, that this is always possible.

It should be noted that the solution of Eq.(\ref{eq11}) does not only
represent the emission and absorption of null dust fluids of 
cosmic strings, but also the
emission and absorption of gravitational waves.
To see this, let us
calculate the corresponding Weyl tensor, which is thought of as
representing the pure gravitational field 
\cite{S1965,W1991}. It can be shown
that it has only three non-vanishing components, $\Psi_{A},\; (A = 0, 2,
4)$, the definitions of them are given by Eq.(\ref{eq9}), 
and each of them is  given by
\bqn
\lb{eq16}
\Psi_{0} &=& e^{a(u)+b(v)}\Psi_{0}^{cs} + \Psi_{0}^{n},\nb\\
\Psi_{2} &=& e^{a(u)+b(v)}\Psi_{2}^{cs},\nb\\
\Psi_{4} &=& e^{a(u)+b(v)}\Psi_{4}^{cs} + \Psi_{4}^{n},
\eqn
where $\Psi_{A}^{cs}$ are given by Eq.(\ref{eq9}), and 
\bqn
\lb{eq17}
\Psi_{0}^{n} &=& - \frac{b'(v)e^{\Omega}}{2\sqrt{2}} 
\left[2(U_{,t} - U_{,r})
- \frac{W_{,t} - W_{,r}}{W} \right],\nb\\
\Psi_{4}^{n} &=& - \frac{a'(u)e^{\Omega}}{2\sqrt{2}} 
\left[2(U_{,t} + U_{,r})
- \frac{W_{,t} + W_{,r}}{W} \right],
\eqn
with $\Omega \equiv 2(U - K) + a(u) + b(v)$.  
The $\Psi^{cs}_{A}$'s represent the gravitational field
of the cosmic string, $\Psi^{n}_{0}$ represents the outgoing
gravitational wave emitted by the string, and $\Psi^{n}_{4}$ the
ingoing gravitational wave absorbed by the string 
\cite{WS1996,W1991}.

\section{DISCUSSIONS}

The back reaction of the gravitational and particle radiation to the
spacetime can be studied by considering the Kretschmann scalar, which
can be written in the form \cite{LW1994}
\bqn
\lb{eq18}
{\cal{R}} &\equiv& R_{\mu\nu\lambda\sigma} R^{\mu\nu\lambda\sigma}\nb\\
&=& e^{2(a + b)}{\cal{R}}^{cs} + 2e^{\Omega}\sigma\left(\rho^{out} +
\rho^{in}\right) + 4e^{2\Omega}\rho^{out}\rho^{in} \nb\\
& & + 16\left[\Psi^{n}_{0}\Psi^{cs}_{4} + 
\Psi^{n}_{4}\Psi^{cs}_{0}\right]
 + 16\Psi^{n}_{0}\Psi^{n}_{4},
\eqn
where ${\cal{R}}^{cs}$ is the corresponding  Kretschmann scalar for the
pure cosmic string, given by
\bqn
\lb{eq19}
{\cal{R}}^{cs}  &=& 16\left[ \Psi^{cs}_{0}\Psi^{cs}_{4} + 
3(\Psi^{cs}_{2})^{2} + (\Phi^{cs}_{02})^{2} + 2(\Phi^{cs}_{11})^{2}
\right.\nb\\
& & \left. +\Phi^{cs}_{00}\Phi^{cs}_{22}  +
6(\Lambda^{cs})^{2}\right],
\eqn
which is regular at the axis $r = 0$, as can be seen from
Eqs. (\ref{eq6}), (\ref{eq8}) and (\ref{eq9}). 
The Kretschmann scalar contains five
terms, each of which has the following physical interpretation
\cite{WS1996,LW1994}: The first term represents the contribution of the
string, and, as shown above, it is always finite 
at the axis. The second term
represents the interaction between the string and the two null dust
fluids, while the third term represents the interaction between the two
null dust fluids. The fourth term represents the gravitational
interaction between the gravitational field of the string and the
outgoing and ingoing gravitational waves, while the last term
represents the interaction between the two outgoing and ingoing
gravitational waves. From Eqs.(\ref{eq6}), 
(\ref{eq13}) and (\ref{eq17}) we find that
\bqn
\lb{eq20}
\rho^{out} &\sim& \frac{b'(t)}{r}, \;\;\;
\rho^{in} \sim \frac{a'(t)}{r}, \nb\\
\Psi^{n}_{0} &\sim& \frac{b'(t)}{r} e^{\Omega(t)},\;\;\;
\Psi^{n}_{4} \sim \frac{a'(t)}{r} e^{\Omega(t)},
\eqn
as $r \rightarrow 0$. Combining Eqs.(\ref{eq18}) and (\ref{eq20}), we
find that the interaction between the cosmic string and the two null
dust fluids and the interaction between the gravitational field of the
string and the two gravitational waves make the Kretschmann scalar
diverge like $r^{-1}$ at the axis for generic choice of the functions
$a(u)$ and $b(v)$, while the interaction between the two null dust
fluids and the interaction between the outgoing and ingoing
gravitational waves make the Kretschmann scalar diverge like $r^{-2}$.

Therefore, it is concluded that the back reaction of emission and
absorption of gravitational and particle radiation of a cosmic string
{\em always} makes the symmetry axis of the string singular. 
The singularity is a scalar one and cannot be removed.  
Our considerations do not refer to any particular solutions, 
and consequently are general.

It should be noted that the existence of the gravitational waves or
null dust fluids in our model may be considered as incidental, 
in the sense that we can not switch off one of them while maintaining 
the other. However, it seems that the conclusion about the formation 
of the spacetime singularities on the axis will not be changed dramatically,
even in the models where they are separable. This can be partially seen 
 from Eqs.(\ref{eq18}). Of course, it would be very interesting to
construct models where they are separable and then study their effects
independently.  
Moreover, instead of taking the results about the formation of spacetime
singularities on the axis as something against
the scenario of the structure and galaxies formations from topological
defects, we would like to consider them as the indication of the 
important role that the back reaction of gravitational and particle
radiation may play, and more realistic models,
such as, strings with finite length and non-cylindrical symmetry, may avoid
the formation of these singularities. 

\section*{Acknowledgments}

The financial assistance from CNPq, FAPERJ and UERJ (AW) is gratefully 
acknowledged.

\end{document}